\documentclass{cernyrep}
\usepackage[T1]{fontenc}
\usepackage[bookmarks, colorlinks=true, linktoc=page, pdftex, linkcolor=black, citecolor=black, urlcolor=blue]{hyperref}

\usepackage{fancyhdr}
\fancyhfoffset{4 mm}
\fancypagestyle{ARTTITLE}{%
\fancyhf{} 
\lhead{\hspace*{0cm} CERN Accelerator School Proceedings ---
{\it Beam Instrumentation} ---  Split,  Croatia, 2025\hfill}
\lfoot{\hspace{3mm} Available online at \url{https://cas.web.cern.ch/previous-schools}}
\rfoot{\thepage\hspace*{3mm}}
 
}

\usepackage{varwidth}
\usepackage{xcolor}

\usepackage{mathtools}
\newcommand{\tvect}[2]{%
	\ensuremath{\Bigl(\negthinspace\begin{smallmatrix}#1\\#2\end{smallmatrix}\Bigr)}}
\begin{document}

\title{Transverse Emittance and Emittance Measurements}
\author{Gero Kube}
\institute{Deutsches Elektronen Synchrotron (DESY), Hamburg, Germany}

\begin{abstract}
	Transverse emittance is one of the central figures of merit for charged-particle beams because it connects the microscopic phase-space distribution to macroscopic accelerator performance. This report introduces the trace-space description of emittance, relates it to the Courant--Snyder formalism and the second-moment beam matrix, and then follows the experimental logic behind common diagnostics. Emphasis is placed on how profile measurements, masks, drifts, quadrupole scans, and transverse deflecting structures convert otherwise inaccessible angular or slice information into measurable beam sizes.
\end{abstract}
\maketitle
\thispagestyle{ARTTITLE}

\section{Introduction}

The transverse emittance of a particle beam describes the area occupied by the beam in transverse phase space. It is a fundamental parameter used to characterize charged-particle beams in accelerators, linking the microscopic phase-space distribution of the particles to the macroscopic performance of the machine.

Together with the Courant–Snyder (Twiss) parameters, the emittance determines the evolution of the beam envelope and, under linear beam transport, forms an invariant quantity that is conserved throughout the accelerator lattice. Consequently, the emittance provides essential information about beam quality, focusing properties, and achievable machine performance. It also plays a central role in accelerator design and operation, entering directly into considerations such as aperture requirements, matching tolerances, synchrotron-radiation brightness, collider luminosity, and the gain process in free-electron lasers.

Because emittance is a statistical property of the beam, several complementary definitions are used in accelerator physics, each emphasizing different aspects or simplifying the analysis for particular applications. Likewise, a broad range of experimental techniques has been developed for emittance measurements, with different methods being better suited to specific beam parameters and accelerator environments. A proper interpretation of emittance measurements therefore requires an understanding of both the underlying definitions and the assumptions and limitations inherent to the corresponding diagnostic techniques.

Experimentally, emittance measurements usually rely on transverse beam-profile diagnostics, with the emittance reconstructed from beam-profile measurements acquired at one or several locations along a beam line. Although these measurements are carried out in geometric space, the corresponding phase-space distribution must be inferred using beam-transport models and reconstruction techniques. This requires a thorough understanding of the relationship between the particle distribution in geometric space and its representation in phase space.

In this report, the concept of transverse emittance and the corresponding measurement techniques are reviewed. The first part summarizes the theoretical foundations required for the interpretation of emittance measurements, including trace space, the Courant–Snyder invariant, transport and beam matrices, statistical beam moments, and normalized emittance. The second part briefly reviews the impact of emittance on beam and machine performance, whereas the third part addresses emittance diagnostics, emphasizing the common physical principle underlying a wide range of measurement techniques: a diagnostic system must either sample the transverse phase space directly or reconstruct the beam matrix from measurements performed under known optical transformations. Further details on these topics can be found in Refs.~\cite{CourantSnyder1958,Wiedemann2015,MintyZimmermann2003}.

\section{Emittance Basics}

\subparagraph{Trace Space and Single-Particle Motion} In classical mechanics, the motion of individual particles is described in a six-dimensional phase space defined by the canonical coordinates ($x, y, z; p_x, p_y, p_z$). If the motion in the three spatial directions can be considered decoupled, the dynamics are commonly analyzed in terms of the projections onto the three orthogonal subspaces rather than in the full six-dimensional phase space. In accelerator physics, where the co-moving coordinate system $(x, y, s)$ is conventionally employed, particle motion is commonly described in trace space rather than in canonical phase space. Under the assumption that the reference momentum $p_0$ is conserved and that $p_s \approx p_0$, the momentum coordinates can be replaced by the corresponding trajectory divergences. For instance, the transverse momentum coordinate $p_x$ is expressed as $p_x \rightarrow x' = p_x/p_s$. This approximation is well justified in the paraxial regime, where the transverse momenta are small compared with the longitudinal momentum.

The transverse dynamics of a single particle in an accelerator can, in one transverse plane, be described as a quasi-harmonic oscillation,
\begin{equation}
	x(s)=\sqrt{\varepsilon\beta(s)}\cos\bigl(\psi(s)+\phi\bigr),
	\label{eq:spx}
\end{equation}
where $\beta(s)$ denotes the beta function, $\psi(s)$ the betatron phase advance, $\phi$ the initial phase, and $\varepsilon$ the~single-particle invariant, commonly referred to as the emittance. Both the oscillation amplitude $a \coloneqq \sqrt{\varepsilon\beta(s)}$, and the phase advance $\psi(s)$ depend on the longitudinal coordinate $s$.

The corresponding trace-space coordinate $x'$ is obtained from the derivative with respect to $s$,
\begin{equation}
	x' = \frac{{\mathrm d}x}{{\mathrm d}s}
	=
	-\sqrt{\frac{\varepsilon}{\beta(s)}}
	\left[
	\alpha(s)\cos\bigl(\psi(s)+\phi\bigr)
	+
	\sin\bigl(\psi(s)+\phi\bigr)
	\right],
	\label{eq:spxp}
\end{equation}
where $\alpha(s) =-\beta'(s)/2$ is the Twiss, or Courant--Snyder, parameter. Analogously, the quantity $b \coloneqq -\sqrt{\varepsilon/\beta(s)}$ defines the characteristic scale of the angular oscillation.

For simplicity, the following discussion considers a particular location $s_0$ in the accelerator lattice corresponding to a beam waist, where $\alpha(s_0) = 0$. Introducing the abbreviation $\theta \coloneqq \psi(s_0)+\phi$, the~phase-space coordinates at $s_0$ reduce to
\begin{equation}
	x(s_0) = a \cos\theta \; \qquad x'(s_0) = b \sin\theta \; .
	\label{eq:spell}
\end{equation}
These expressions correspond to the standard parametric representation of an ellipse with semi-axes $a$~and $b$ in the trace space $(x,x')$. Thus, at a given longitudinal position $s$, the transverse position and angle of an individual particle evolve over time along an elliptical trajectory in phase space. From classical geometry, the area $A$ of an ellipse is given by $A = \pi\,|a|\,|b|$. Substituting the definitions of $a$ and $b$ yields
\begin{equation}
	A = \pi\,\varepsilon.
	\label{eq:spemit}
\end{equation}
This relation corresponds to the conventional definition of the \emph{geometric emittance}: it is equal to the~phase-space area enclosed by the ellipse $A$ divided by $\pi$. For a real particle beam, this concept is generalized to a statistical quantity describing the ensemble distribution in transverse phase space \cite{CourantSnyder1958}.

\subparagraph{Dispersion as Offset in Trace Space} Using the ellipse parametrization in Eq.~(\ref{eq:spell}), it is mathematically evident that an offset $x_0$, respectively $x'_0$, shifts the ellipse along the horizontal or vertical axis in trace space, while its shape and orientation remain unchanged.

The physical interpretation of such an offset is the dispersion, which describes the dependence of the transverse particle trajectory on the particle momentum. For a relative momentum deviation $\delta=\Delta p/p$, the dispersion terms shift the trace-space coordinates of off-momentum particles according to
\begin{equation}
	x = x_\beta + D\delta, \qquad x' = x'_\beta + D'\delta \; .
	\label{eq:DispContrib}
\end{equation}
For an individual particle, this shift neither changes the area nor the orientation of the betatron ellipse. However, for a beam with finite momentum spread, it increases the observed beam size and, correspondingly, the beam divergence. Therefore, emittance measurements are most reliable in regions where the~dispersion is either negligible or precisely known and properly corrected for.

\subparagraph{Courant--Snyder Parameters and Beam Matrix} Away from a beam waist, the phase-space ellipse is generally tilted. Nevertheless, the preceding conclusions remain valid: at an arbitrary longitudinal position $s$, the particle trajectory follows an ellipse in phase space, and the geometric emittance is given by Eq.~(\ref{eq:spemit}). In the general case, the shape and orientation of the emittance ellipse are characterized by the~Courant–Snyder parameters $\alpha$, $\beta$, and $\gamma$, related by
\begin{equation}
	\gamma(s)=\frac{1+\alpha^2(s)}{\beta(s)}.
\end{equation}
\begin{figure}[!b]
	\centering
	\includegraphics[width=0.72\textwidth]{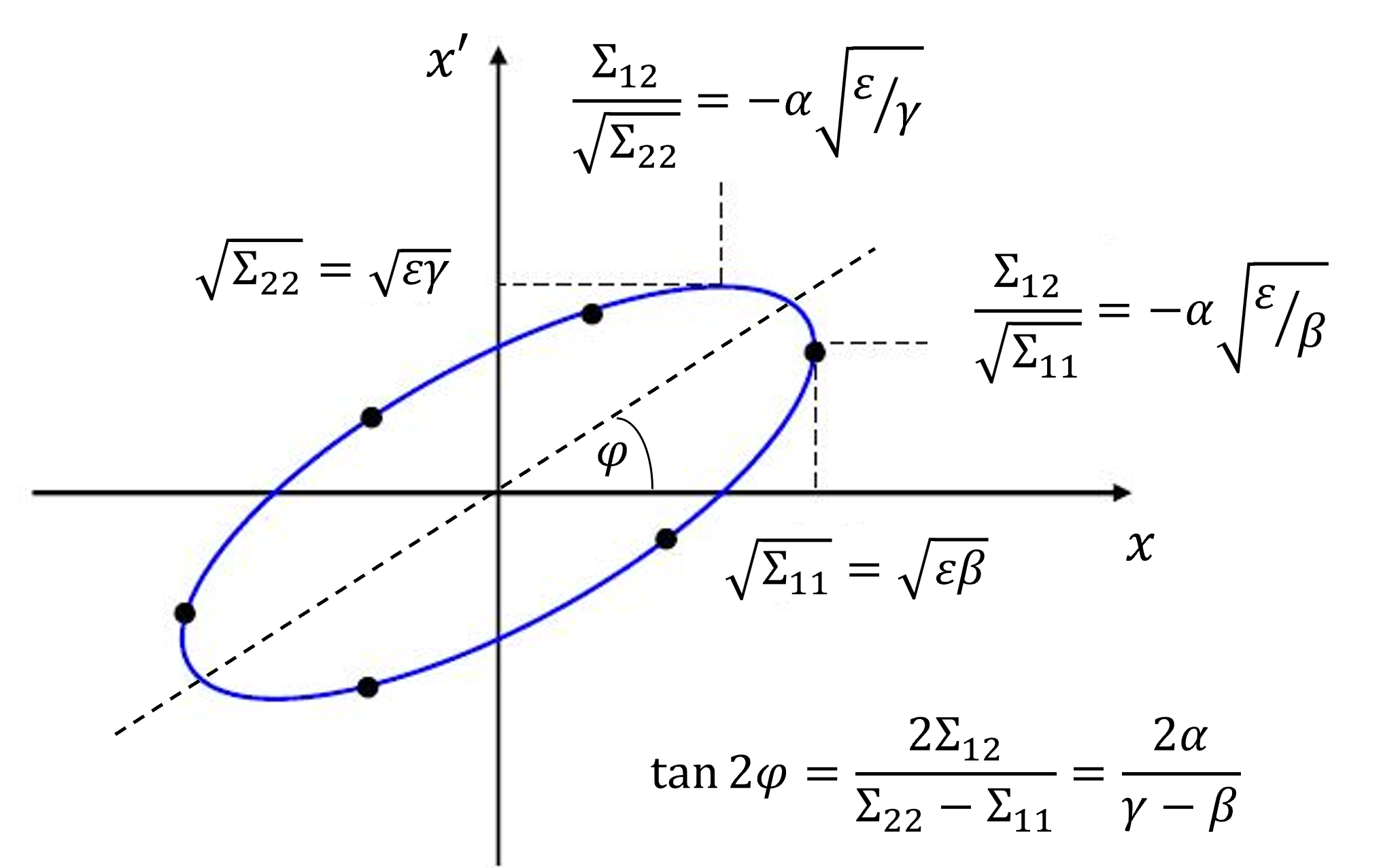}
	\caption{General trace-space ellipse, described either by the beam matrix formalism or by the equivalent Courant–Snyder parametrization.}
	\label{fig:beam_matrix}
\end{figure}
Using this parametrization, the ellipse equation can be written as
\begin{equation}
	\varepsilon = \gamma x^2 + 2\alpha x x' + \beta x'^2 \; ,
	\label{eq:cs_ellipse1}
\end{equation}
which may be expressed in matrix form as
\begin{equation}
	\vec{x}^{\mathsf{T}} \, \Sigma^{-1} \vec{x} = 1, 
	\label{eq:cs_ellipse}
\end{equation}
where $\vec{x} = \tvect{x}{x'}$ denotes the trace-space vector. The \emph{beam matrix} $\Sigma$, introduced in Eq.~(\ref{eq:cs_ellipse}), is defined as
\begin{equation}
	\Sigma =
	\begin{pmatrix}
		\Sigma_{11} & \Sigma_{12} \\
		\Sigma_{12} & \Sigma_{22}
	\end{pmatrix}
	=
	\varepsilon
	\begin{pmatrix}
		\beta & -\alpha \\
		-\alpha & \gamma
	\end{pmatrix}.
	\label{eq:sigma}
\end{equation}
Figure~\ref{fig:beam_matrix} illustrates the trace-space ellipse together with the beam matrix representation and the corresponding Courant–Snyder parametrization.

\begin{figure}[!h]
	\centering
	\includegraphics[width=0.85\textwidth]{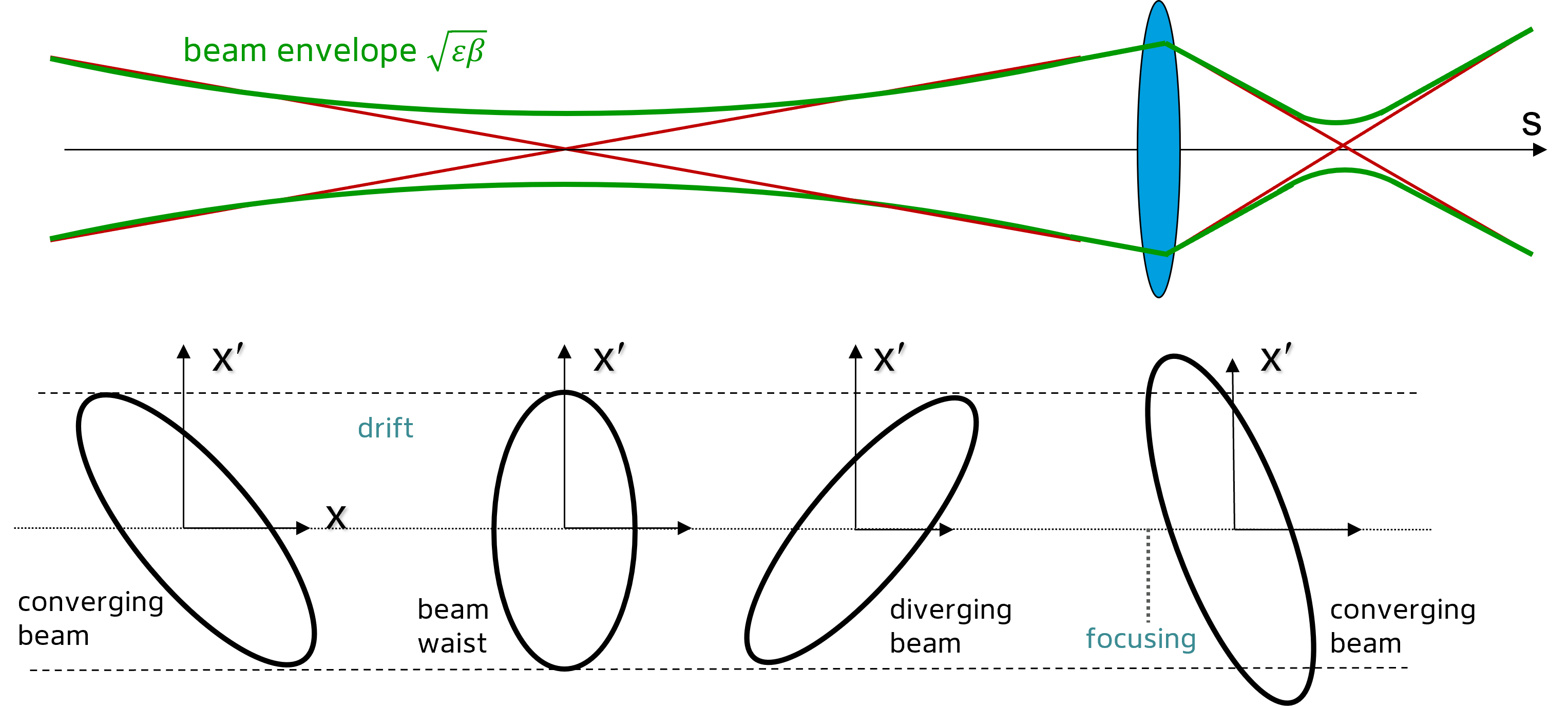}
	\caption{Trace space evolution in drift spaces and quadrupoles. In drift spaces the ellipses are sheared horizontally, whereas in quadrupoles they are sheared vertically.}
	\label{fig:space_evol}
\end{figure}

\subparagraph{Trace Space Evolution and Particle Transport} The upper plot in \Fref{fig:space_evol} shows the evolution of the beam through a drift section around a beam waist, followed by a focusing element (quadrupole). The~green lines denote the beam envelope formed by an ensemble of particles, whereas the red line highlights the trajectory of a representative single particle. The lower plot depicts the corresponding evolution of a single particle ellipse in trace-space. It can be seen that passage through a drift space leaves the~single-particle divergence coordinate $x'$ unchanged, while  emittance projection onto the transverse coordinate $x$ evolves. Consequently, the ellipse is progressively sheared horizontally as the beam propagates through the drift section.
Considering the quadrupole action in the thin-lens approximation, the beam size at the quadrupole location (corresponding to the ellipse projection onto the transverse coordinate $x$) remains unchanged due to the vanishing longitudinal extent of the element, whereas the divergence is modified and may even change sign. Consequently, the action of the quadrupole corresponds to a vertical shearing of the trace-space ellipse.

According to Liouville’s theorem \cite{Liouville1838}, the particle density in phase space remains constant under the action of conservative forces. In the language of accelerator physics, this implies that the beam emittance is conserved under linear forces, such as those generated by quadrupoles and drift spaces, which are the typical elements in storage rings and beam transfer lines \cite{Wiedemann2015}. Consequently, although the~shape and orientation of the trace-space ellipse evolve during beam transport through the accelerator lattice, its area remains invariant.

\begin{figure}[!t]
	\centering
	\includegraphics[width=0.7\textwidth]{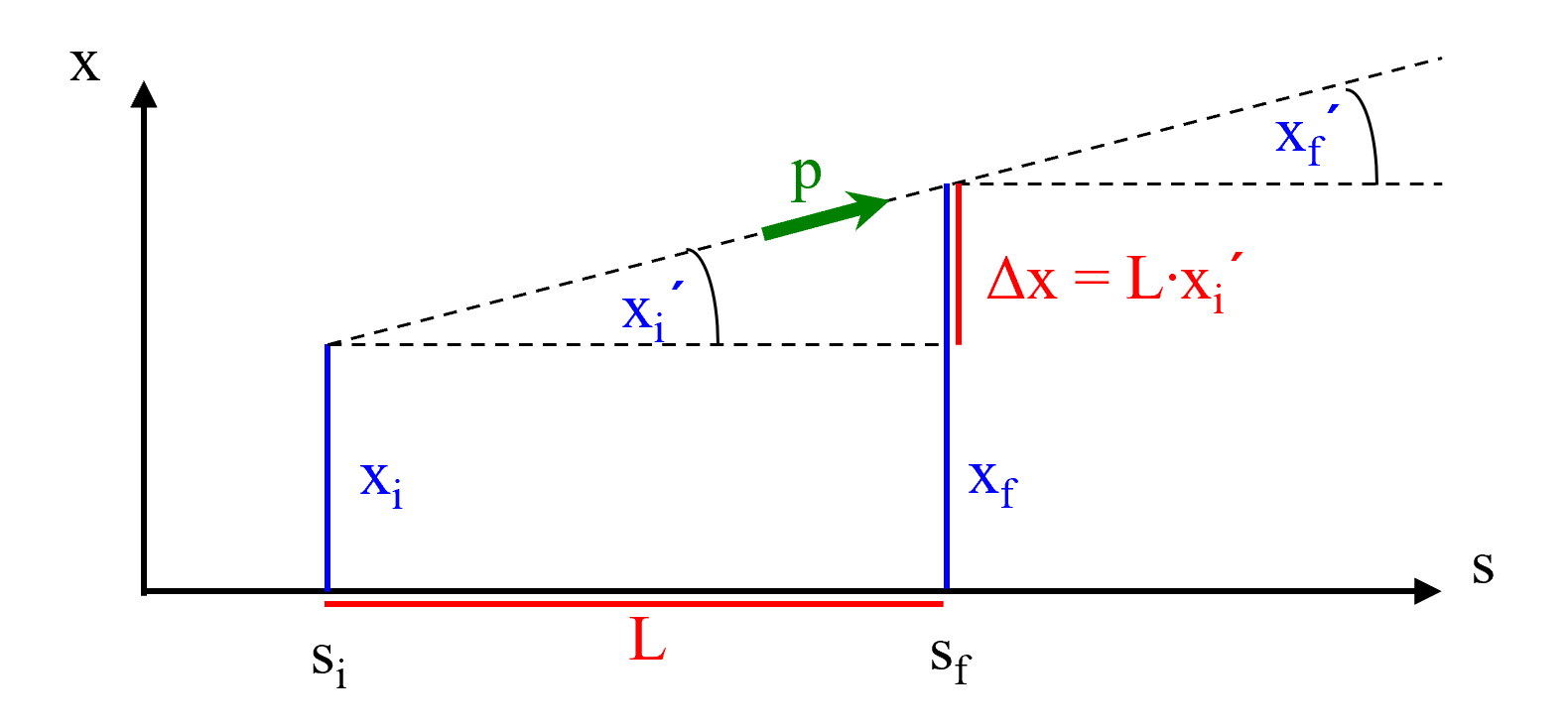}
	\caption{Geometrical interpretation of particle propagation. The particle $p$ propagates through a drift space of length $L$ from the initial position $s_i$ in the accelerator lattice to the final position $s_f$.}
	\label{fig:drift}
\end{figure}

In analogy to the formalism developed for classical paraxial optics, the propagation of charged particles through accelerator lattice elements is commonly described using the well-established transfer matrix formalism. For a single particle propagating through the lattice, the trace-space vector $\tvect{x}{x'}$ is transformed from its initial location ($i$) to its final one ($f$) according to
\begin{equation}
	\begin{pmatrix}x\\x'\end{pmatrix}_{f}
	=
	R
	\begin{pmatrix}x\\x'\end{pmatrix}_{i},
	\qquad
	R = \begin{pmatrix}R_{11}&R_{12}\\R_{21}&R_{22}\end{pmatrix} \; .
\end{equation}
The matrix elements of $R$ depend on the optical elements located between the two positions. For the case of a simple drift space, the formalism allows a straightforward geometric interpretation, see \Fref{fig:drift}.
As can be seen from this illustration, the divergence coordinate $x'$ remains unchanged during propagation through the drift space, whereas the transverse position at $s_f$ acquires an additional offset of $\Delta x = L \cdot x'_i$:
\begin{align*}
	x_f & = 1 \cdot x_i + L \cdot x'_i \\
	x'_f & = 0 \cdot x_i + 1 \cdot x'_i
\end{align*}
This set of equations can be compactly expressed in matrix form as
\begin{equation}
	\begin{pmatrix}x\\x'\end{pmatrix}_{f} = \underbrace{\begin{pmatrix}1&L\\0&1\end{pmatrix}}_{R_d} \, \begin{pmatrix}x\\x'\end{pmatrix}_{i}
	\label{eq:drift}
\end{equation}
with the drift transfer matrix $R_d$.

Instead of applying the single-particle transformation, the Courant–Snyder parameters can be propagated directly through the accelerator lattice using a transfer matrix constructed from the elements of the single-particle transfer matrix $R$:
\begin{equation}
	\begin{pmatrix} \beta \\ \alpha \\ \gamma \end{pmatrix}_{f} = 
	\begin{pmatrix}
		R_{11}^2 & -2 R_{11} R_{12} & R_{12}^2 \\
		-R_{11} R_{21} & R_{11} R_{22} + R_{21} R_{12} & -R_{12} R_{22} \\
		R_{21}^2 & -2 R_{21} R_{22} & R_{22}^2
	\end{pmatrix}
	\cdot
	\begin{pmatrix} \beta \\ \alpha \\ \gamma \end{pmatrix}_{i} \; .
	\label{eq:transferm}
\end{equation}
As a third approach, the beam matrix introduced in Eq.~(\ref{eq:sigma}) can be propagated through the accelerator lattice using the single-particle transfer matrix $R$. In this case, the transformation is given by
\begin{equation}
	\Sigma_f = R \cdot \Sigma_i \cdot R^{\mathsf{T}} .
	\label{eq:sigma_transport}
\end{equation}
As will be shown in the following sections, the latter transformation forms the basis of many commonly used beam emittance measurement techniques.

\subparagraph{Multi-Particle and Statistical Emittance} Up to this point, the discussion has been restricted to single-particle dynamics. In a real accelerator, however, a particle beam consists of a large ensemble of individual particles rather than a single trajectory, with each particle following its own path through trace space. To describe such a system, the emittance must be treated statistically in terms of the collective particle distribution in trace space $\rho(x,x')$. The transverse beam emittance then provides a quantitative measure of the trace-space area occupied by the beam, thereby characterizing both its spatial extent and angular divergence.

For an ensemble of particles, the trace-space distribution can be characterized by its second-order moments, 
\[
\langle x^2 \rangle = \frac{\int_{-\infty}^{+\infty}\mathrm{d}x\mathrm{d}x'(x-\mu)^2 \rho(x,x')}{\int_{-\infty}^{+\infty}\mathrm{d}x\mathrm{d}x' \rho(x,x')},
\]
where $\mu$ denotes the first-order central moment. Using this notation, the beam matrix formalism Eq.~(\ref{eq:sigma}) can be written in a more generalized form as
\begin{equation}
	\Sigma =
	\begin{pmatrix}
		\Sigma_{11} & \Sigma_{12} \\
		\Sigma_{12} & \Sigma_{22}
	\end{pmatrix}
	=
	\begin{pmatrix}
		\langle x^2\rangle & \langle x x'\rangle \\
		\langle x x'\rangle & \langle x'^2\rangle
	\end{pmatrix}
	=
	\varepsilon
	\begin{pmatrix}
		\beta & -\alpha \\
		-\alpha & \gamma
	\end{pmatrix}.
	\label{eq:sigma1}
\end{equation}
The beam matrix plays a central role because the beam emittance can be directly determined from its elements: the \emph{rms emittance} is equal to the square root of the determinant of the matrix,
\begin{equation}
	\varepsilon_{\rm rms}=\sqrt{\det\Sigma}
	=\sqrt{\Sigma_{11}\Sigma_{22}-\Sigma_{12}^2} \; .
	\label{eq:rms_emit}
\end{equation}
Equation~(\ref{eq:rms_emit}) directly provides a practical method for determining the beam emittance, which is widely used in beam diagnostics. The task therefore reduces to determining the elements of the beam matrix, from which the emittance can subsequently be obtained using Eq.~(\ref{eq:rms_emit}). This approach will be discussed in later sections in the context of emittance measurement techniques.

When comparing emittance values reported for different accelerators, it should be noted that various conventions are used throughout the accelerator community. Since most particle beams exhibit an~approximately Gaussian, or bell-shaped, transverse distribution, the standard convention for electron beams is to quote the emittance in terms of the $1\sigma$ width of the Gaussian distribution. In a proton machine the~emittance boundary is conventionally chosen to include about 90\% (strictly 87\%) of a Gaussian beam at $2\sigma$ \cite{Wilson2001}.

\subparagraph{Adiabatic Damping and Normalized Emittance} So far, the emittance has been treated as an~invariant quantity in trace space. This assumption is valid as long as particle acceleration is neglected. Recalling the definition of the trajectory divergence, $x' = p_x/p_s$, acceleration increases the longitudinal momentum component $p_s$, while the transverse momentum $p_x$ remains approximately unchanged. Consequently, the divergence decreases with increasing beam energy, leading to a reduction of the emittance. This effect is referred to as transverse \emph{adiabatic damping}.

For a deeper understanding, it should be recalled that Liouville’s theorem states the conservation of particle density in canonical phase space rather than in trace space. Consequently, the emittance should formally be expressed in terms of Hamilton’s canonical phase-space coordinates $(q, p)$ and subsequently related to the corresponding trace-space variables. Here, q denotes a general transverse coordinate (previously represented by $x$), while $p$ denotes the corresponding conjugate momentum.

According to Liouville, the phase-space area enclosed by the contour, $A = \int p\,\mathrm{d}q$, is conserved. The conjugate momentum can be rewritten as
\[
p = m\,v = \gamma m_0 \frac{\mathrm{d}q}{\mathrm{d}t} = \gamma m_0 \frac{\mathrm{d}q}{\mathrm{d}s}\frac{\mathrm{d}s}{\mathrm{d}t} =  m_0 c \; \beta \gamma  q' \; . 
\]
Substituting this expression into the phase-space integral yields
\[
A = \int p\,\mathrm{d}q = m_0 c \; \beta \gamma \underbrace{\int\,\mathrm{d}q \, q'}_{\varepsilon} \; .
\]
Since the phase-space area $A$ is conserved and both $m_0$ and $c$ are constants, the remaining factor must also remain invariant. This quantity is referred to as the \emph{normalized emittance} $\varepsilon_N$ and is defined as
\begin{equation}
	\varepsilon_N = \beta \gamma \, \varepsilon \; .
	\label{eq:emit_n}
\end{equation}

\section{Emittance Impact}

The beam emittance is one of the most important parameters in accelerator physics because it quantifies the beam quality in transverse phase space. In addition, it directly determines the transverse beam size within the accelerator according to
\begin{equation}
	\sigma = \sqrt{\beta\varepsilon} \, ,
	\label{eq:BeamSize}
\end{equation}
where $\sigma$ and $ \varepsilon$ denote the rms beam size and the rms emittance, respectively. This relation follows directly from Eq.~(\ref{eq:sigma1}), taking into account that the rms beam size is given by $\sigma = \sqrt{\langle x^2\rangle} = \sqrt{\Sigma_{11}}$. In the~following, several key advantages associated with low beam emittance are summarized.

\subparagraph{Aperture Limitation}
The most immediate limitation is imposed by the available aperture, which must provide sufficient clearance between the beam and the vacuum chamber. For a matched beam with negligible dispersion, the transverse beam size is given by Eq.~(\ref{eq:BeamSize}). In this sense, the beam emittance defines the characteristic scale of the beam within the accelerator and strongly influences how efficiently the machine can be operated.

In electron accelerators, a 2$\sigma$ boundary would lie far too close to the beam core. An aperture restriction placed at such a distance would rapidly intercept particles as they diffuse into the beam tails due to quantum excitation and radiation damping. The required physical aperture therefore depends on the~desired beam lifetime, but for electron machines it typically corresponds to a distance of approximately 6$\sigma$ to 10$\sigma$  from the beam center \cite{Wilson2001}.

\subparagraph{Mismatch and Optics Corrections}
A beam may also possess an acceptable emittance while still having incorrect Twiss parameters. In such a case, the ellipse area is preserved, but its orientation and aspect ratio are mismatched with respect to the design optics. As a consequence, the beam envelope oscillates along the lattice, which can lead to beam losses, reduced injection efficiency, emittance growth after transport through nonlinear elements, or degraded performance.
\begin{figure}[!h]
	\centering
	\includegraphics[width=0.7\textwidth]{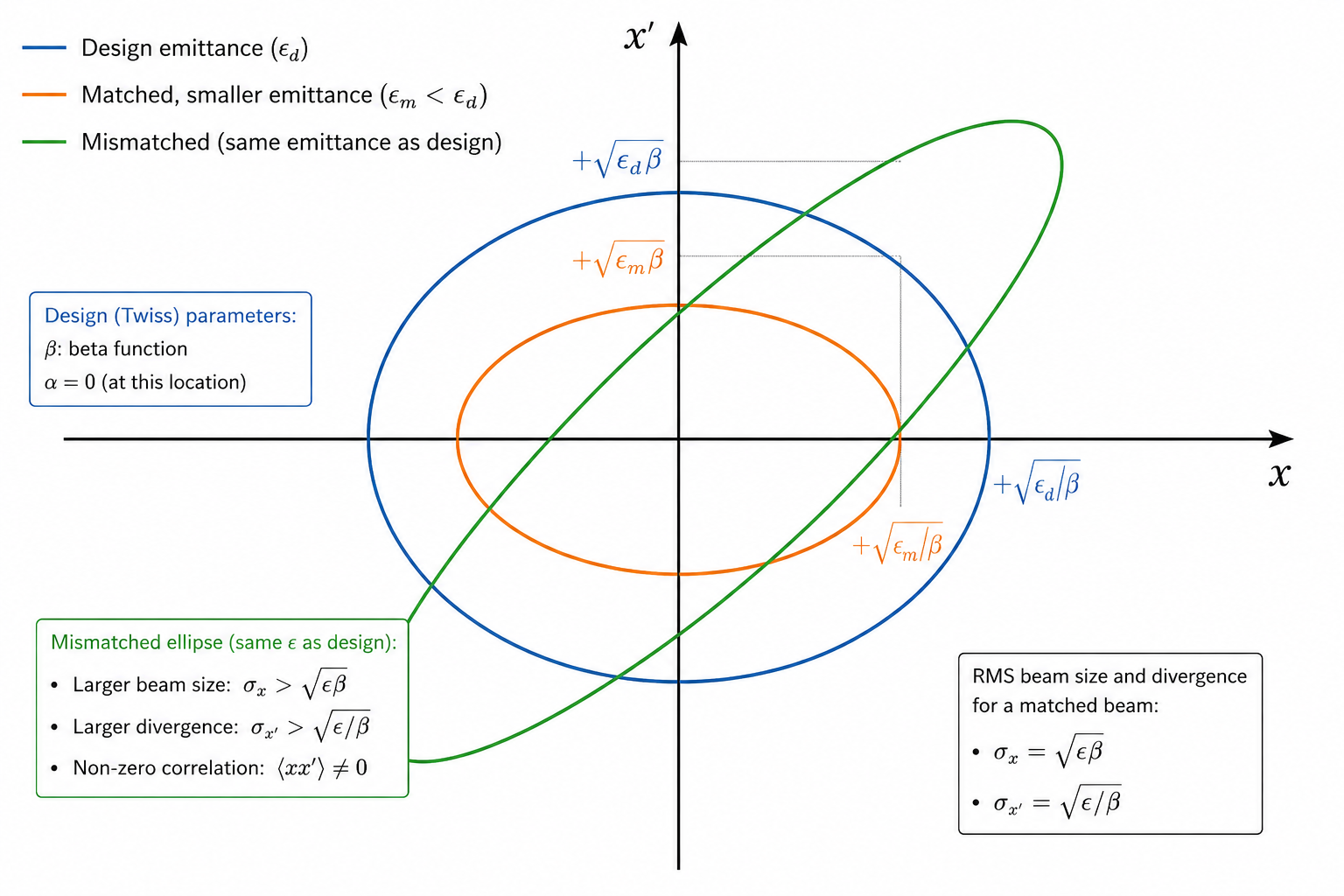}
	\caption{Design emittance ellipse (blue) together with a matched ellipse of smaller trace space area (orange). The~tilted ellipse (green) represents a beam that is mismatched with respect to the design optics, despite having the~same trace-space area.}
	\label{fig:mismatch}
\end{figure}

\subparagraph{Accelerator Performance - Light Sources and Colliders}
In synchrotron light sources, accelerator performance is commonly evaluated in terms of \emph{spectral brightness} $\mathcal{B}$, which is defined as
\[
\mathcal{B} = \frac{\mathrm{number~of~photons}}{\mathrm{[s]}\,\mathrm{[mm^2]}\,\mathrm{[mrad^2]}\,\mathrm{[0.1\%~bandwidth]}} \; .
\]
Light sources users require a beam with high brightness, corresponding to a high photon flux delivered onto their sample emitted from a source with small transverse size and low divergence, within a narrow spectral bandwidth. Relating the brightness to the electron beam parameters, it scales approximately as
\[
\mathcal{B} \propto \frac{N_{\gamma}}{\sigma_x\sigma_{x'}\sigma_y\sigma_{y'}} \propto \frac{I}{\varepsilon_x\varepsilon_y} \; ,
\]
where $N_{\gamma}$ is the photon flux, $I$ the electron beam current, and $\varepsilon_{x,y}$ the transverse emittances. Hence, achieving high brightness requires electron beams with very small transverse emittances.

Moreover, in X-ray free-electron lasers, the transverse emittance determines the overlap between the electron beam and the radiation field, thereby constraining the achievable wavelength according to
\[
\varepsilon_{x,y} < \frac{\lambda}{4\,\pi} \; .
\]
 
For colliders, the analogous performance metric is luminosity $\mathcal{L}$. It is a relativistic invariant proportionality factor between the reaction cross section $\sigma$ (which is a property of interaction under investigation) and the number of interactions per second $\dot{N}$, 
\[
\dot{N} = \mathcal{L} \, \sigma \; .
\]
An experimentalist at a collider aims to study a specific reaction channel with high statistical significance. This requires high event rates and, consequently, a collider with high luminosity. The luminosity is inversely proportional to the transverse beam sizes at the interaction point. For two identical colliding beams, it scales as
\[
\mathcal{L} \propto \frac{1}{\sigma_x\sigma_y} \propto \frac{1}{\sqrt{\varepsilon_x\varepsilon_y}} \; ,
\]
showing that high luminosity requires beams with small transverse emittances.

\section{Emittance Measurements}

A measurement of the transverse emittance requires determining the projected area occupied by the~beam in transverse trace space. However, trace space is not directly accessible to beam diagnostics, since beam monitors perform measurements in real space, described by the coordinates ($x, y, z$). The quantities accessible to beam diagnostics are the projections onto the coordinate axes, as illustrated in \Fref{fig:beam_matrix}, which are given by
\begin{align}
	\mathrm{beam~size} \qquad & \sigma = \sqrt{\Sigma_{11}} = \sqrt{\langle x^2\rangle} = \sqrt{\varepsilon\beta} \, , \label{eq:Bsize} \\
	\mathrm{beam~divergence} \qquad & \sigma' = \sqrt{\Sigma_{22}} = \sqrt{\langle x'^2\rangle} = \sqrt{\varepsilon\gamma} \, . \label{eq:Bdivergence}
\end{align}
Since direct beam-divergence measurements are only rarely employed in practice, the following discussion is restricted to beam-profile measurements. In this context, the specific type of beam profile monitor used is of secondary importance. Further details on beam-profile monitoring techniques can be found in the dedicated report in these proceedings \cite{Torino2026}.

In general, two main schemes for transverse emittance measurements in linacs and transfer lines can be distinguished, which will be described in the following:
\begin{description}
	\item[trace-space mapping,] in which (infinitesimal) beam elements are sampled in trace space, while the~position information $x_D$ measured in the detector plane is converted into the divergence coordinate $x'$ after a drift section;
	\item[beam matrix based measurements,] in which the individual elements of the beam matrix are determined and the emittance is subsequently calculated according to Eq.~(\ref{eq:rms_emit}).
\end{description}
Emittance diagnostics in a storage ring will be discussed in a separate section of this chapter.

\subsection{Trace-Space Mapping}
The principle of trace-space mapping is explained through the slit-scan method, which is described below.

\begin{figure}[!h]
	\centering
	\includegraphics[width=0.98\textwidth]{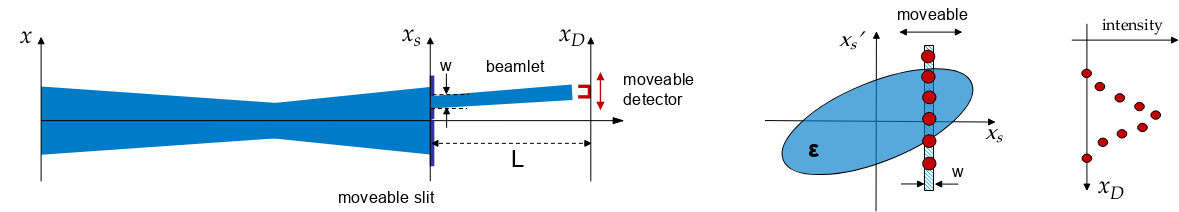}
	\caption{Schematic of the slit-scan method: A narrow slit selects a beamlet at position $x_s$, while a moveable intensity monitor in the detector plane is used to reconstruct the corresponding divergence coordinates $x'_s$.}
	\label{fig:SlitScan}
\end{figure}
\subparagraph{Slit Scan Method} 
This method, illustrated in \Fref{fig:SlitScan}, represents the most direct implementation of trace-space mapping. The goal is to determine the emittance in the plane of the slit mask, denoted by the coordinate $x_s$. A narrow slit of width $w$ selects a beamlet at position $x_s$ in trace space. After traversing a drift of length $L$, the beamlet reaches the detector plane. An intensity monitor (for example, a Faraday cup) is scanned in discrete steps along the detector coordinate $x_D$, recording the corresponding intensity profile. From the slit location $x_s$ and the detector postion $x_D$, the divergence coordinate $x'_s$ is then obtained using the drift transport matrix, yielding
\begin{equation}
	\begin{pmatrix}x_D\\x'_D\end{pmatrix}
	=
	\begin{pmatrix}1&L\\0&1\end{pmatrix}
	\begin{pmatrix}x_s\\x'_s\end{pmatrix},
	\qquad
	x'_s=\frac{x_D-x_s}{L}.
\end{equation}
By moving the slit, the distribution is scanned in the coordinate $x_s$, while the downstream intensity profile measured in the detector plane provides the angular distribution of the selected slice. This method is intuitive and robust; however, it is destructive and relatively slow, since the trace space must be reconstructed from many individual measurements.

Moreover, the mask design involves an important compromise. On the one hand, the mask thickness must be sufficient to stop particles outside the slit opening. On the other hand, excessively thick masks should be avoided, as they can form a channel in which particles undergo scattering within the~slit, thereby distorting the measured phase-space distribution.

\begin{figure}[!t]
	\centering
	\includegraphics[width=0.7\textwidth]{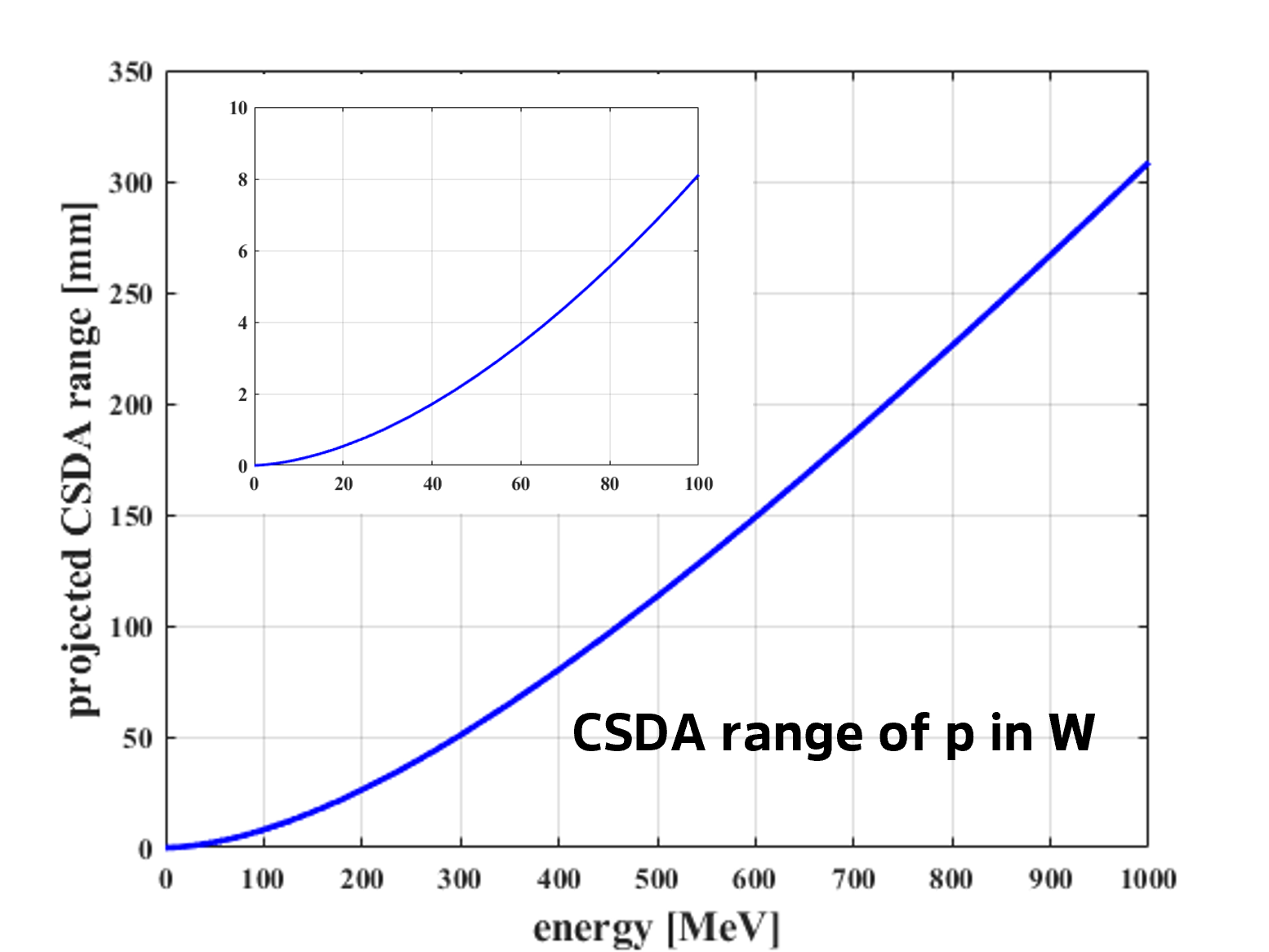}
	\caption{Calculated stopping range of protons in tungsten within the Continuous Slowing Down Approximation (CSDA) as a function of beam energy.}
	\label{fig:slit_range}
\end{figure}

From the requirement that all particles outside the slit opening must be completely blocked, it is evident that this method is only applicable at relatively low particle energies. To illustrate this limitation, \Fref{fig:slit_range} shows the stopping range of protons in tungsten, projected onto the incident beam axis and calculated within the Continuous Slowing Down Approximation (CSDA) as a function of the proton kinetic energy. Details of the CSDA formalism can be found, for example, in Ref.~\cite{Kube2026}. Assuming a practical upper limit of approximately $1\,\mathrm{cm}$ for the mask thickness, it can be concluded that the slit-scan method is applicable only to proton beams with kinetic energies of roughly $T_{kin} \lesssim 100\,\mathrm{MeV}$. Beyond this energy, the required absorber thickness becomes impractically large.

The spatial resolution of this method in $x$-direction is determined by the slit width $w$. Reducing the slit width improves the spatial resolution; however, it also increases the influence of multiple scattering and decreases the transmitted beam intensity, thereby requiring longer acquisition times to achieve sufficient statistical accuracy. The angular resolution is limited by the detector scan step size, $\Delta x_{scan}$, and the detector aperture, $\Delta x_{det}$, yielding approximately
\begin{equation}
	\Delta x' \simeq \frac{\Delta x_{\rm scan}+\Delta x_{\rm det}}{L} \; .
\end{equation}
From this, it can be concluded that the method becomes particularly challenging for beams with small transverse sizes and low angular divergence. Besides the restriction to low particle-beam energies, a~major disadvantage of the slit-scan method is its time-consuming nature, as it requires scanning in both trace-space dimensions. The total number of measurements that must be recorded is therefore given by $N_x \times N_{x'}$, where $N_x$ and $N_{x'}$ denote the numbers of sampling points in the spatial and angular coordinates, respectively. Therefore, the following paragraph presents extensions of the slit-scan method that reduce the measurement time and enable more efficient data acquisition.

\subparagraph{Slit Grid, Multi-Slit, and Pepper Pot} 
\begin{figure}[!b]
	\centering
	\includegraphics[width=0.98\textwidth]{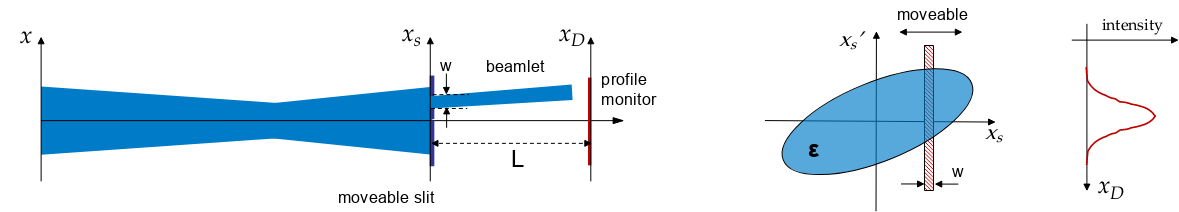}
	\caption{As in the slit-scan method, a slit selects a narrow slice of the beam at position $x_s$. However, instead of scanning an intensity detector along the $x_D$ coordinate, a spatially resolving detector is employed to measure the~beamlet profile in the detector plane.}
	\label{fig:SlitGrid}
\end{figure}
A direct extension of the slit-scan method is the slit-grid-based trace-space mapping technique, which is schematically illustrated in \Fref{fig:SlitGrid}. In this approach, the scanning intensity detector in the $x_D$ plane is replaced by a spatially resolving detector, allowing the angular distribution of the selected beamlet to be recorded in a single measurement. For hadron beams, secondary-electron-emission (SEM) grids are commonly used as spatially resolving detectors; consequently, the technique is often referred to simply as the \emph{slit-grid method}. As a result, only the slit mask has to be scanned along the $x_s$ coordinate, reducing the total number of required measurements to just $N_x$.

\begin{figure}[!h]
	\centering
	\includegraphics[width=0.98\textwidth]{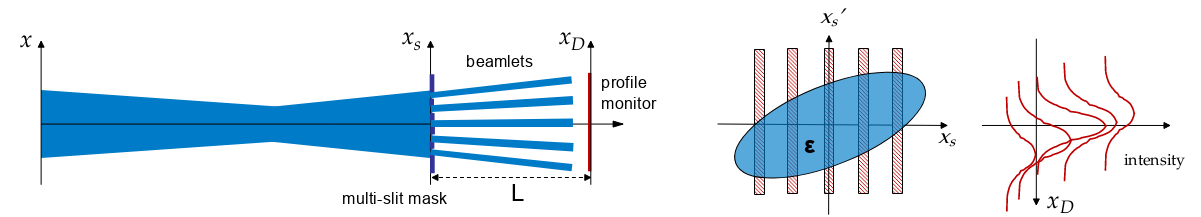}
	\caption{Multi-slit mask concept. Several beamlets are produced at once, eliminating slit scanning.}
	\label{fig:multislit}
\end{figure}

To further reduce the number of required measurements, a multi-slit mask can be placed in the $x_s$ plane to sample multiple positions simultaneously, as illustrated in \Fref{fig:multislit}. In this way, several beamlets are generated in parallel, allowing the trace-space distribution to be reconstructed from a single measurement. Provided that the individual beamlets remain well separated at the spatially resolving detector, this technique enables single-shot emittance measurements in one transverse plane.
A practical implementation of this technique is the emittance meter operated at the SPARC-X facility (INFN Frascati, Italy) \cite{Alesini2007,Cianchi2008}. The device employs a multi-slit mask in combination with a spatially resolving detector, enabling single-shot measurements of the transverse emittance.

\begin{figure}[!b]
	\centering
	\includegraphics[width=0.95\textwidth]{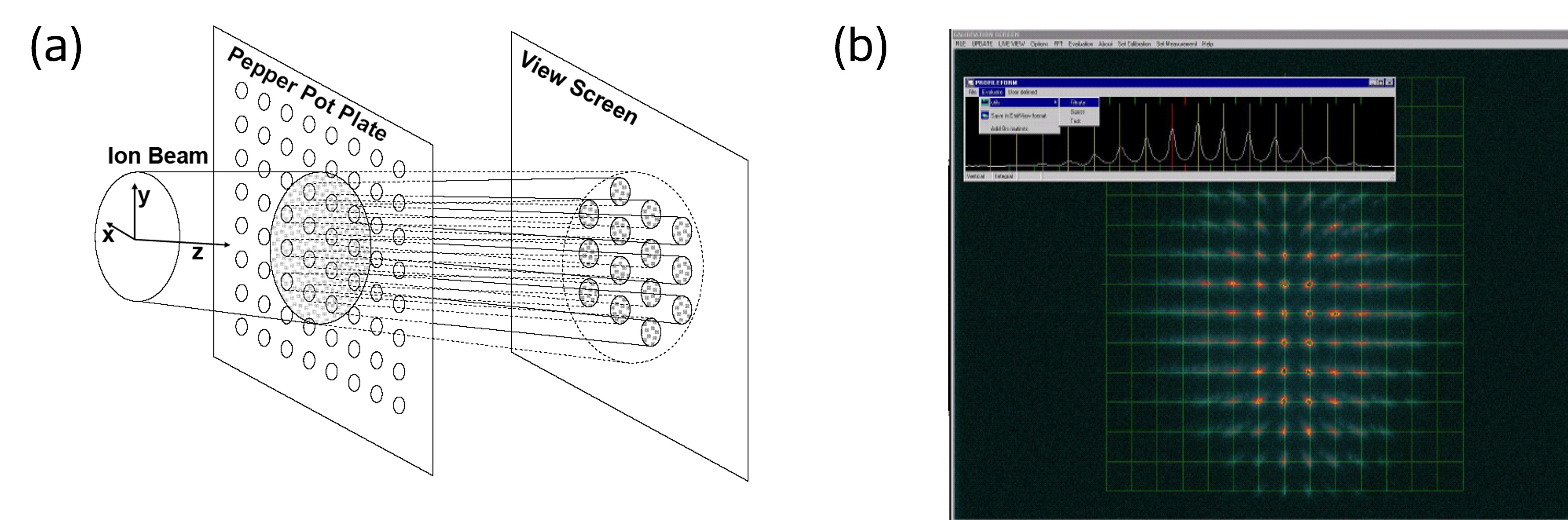}
	\caption{(a) Schematic layout of a pepper-pot setup for simultaneous emittance measurements in both transverse planes. (b) Example of a pepper-pot measurement. Figure adapted from Ref.~\cite{Forck2024}.}
	\label{fig:PepperPot}
\end{figure}
A pepper-pot mask extends the trace-space mapping principle to both transverse planes simultaneously. As illustrated in \Fref{fig:PepperPot}(a), an array of holes generates multiple beamlets whose downstream images encode both the horizontal and vertical position and angular distributions. The mask geometry must satisfy several design criteria. First, the hole spacing should be significantly smaller than the beam size to ensure that the transverse beam profile is sampled by a sufficiently large number of beamlets. Second, the beamlet image size at the detector should be considerably larger than the hole diameter, such that the measured distribution is dominated by the beam divergence rather than the hole geometry. Finally, the hole spacing must be large enough to prevent overlap of neighboring beamlet images at the~detector plane. As an example, \Fref{fig:PepperPot}(b) shows a pepper-pot image obtained with an Ar$^{1+}$ ion beam at an energy of $1.4\,\mathrm{MeV/u}$ \cite{Forck2024}.

While the underlying concept of pepper-pot diagnostics clearly originates from trace-space mapping, the data analysis is essentially based on the calculation of weighted moments and is therefore closely related to beam-matrix-based measurement techniques. For the interested reader, a rigorous treatment of slit-scan and pepper-pot emittance measurements is provided in Ref.~\cite{Zhang1996}.

\subsection{Beam-Matrix Measurements}
\label{sec:BMschemes}

The beam-matrix-based methods exploit Eq.~(\ref{eq:rms_emit}) together with the beam matrix formalism introduced in Eq.~(\ref{eq:sigma1}). To determine the emittance, the three independent matrix elements $\Sigma_{11}$, $\Sigma_{22}$, and $\Sigma_{12}=\Sigma_{21}$ must be determined at the same location $s$. While $\Sigma_{11}$ is directly related to the beam size and can therefore be measured straightforwardly, $\Sigma_{22}$ can, in principle, be obtained from a measurement of the~beam divergence. The correlation term $\Sigma_{12}=\Sigma_{21}$, however, is not directly accessible through beam diagnostics. Consequently, additional information is required to reconstruct the complete beam matrix and thereby determine the emittance.

To this end, beam-matrix-based methods exploit the known transformation properties of the beam matrix $\Sigma$ according to Eq.~(\ref{eq:sigma_transport}). Rather than determining all beam matrix elements at a single accelerator location, the unknown matrix can be reconstructed from a set of at least three matrix element measurements performed under different conditions. According to Eq~(\ref{eq:sigma_transport}), the relation between the corresponding beam matrices is given by
\[
\Sigma(n) = R \cdot \Sigma(0) \cdot R^{\mathsf{T}} \; ,
\]
where $\Sigma(0)$ denotes the beam matrix in a reference state and $\Sigma(n)$ the beam matrix corresponding to the~$n$-th measurement configuration. The transport matrix $R$ describes the beam transport between the~reference state and the measurement state and is assumed to be known from the accelerator optics. Since beam-profile measurements (determination of $\sigma$) provide the most direct access to the beam matrix element
\[
\Sigma_{11} = \varepsilon \beta = \sigma^2 \; ,
\]
the following discussion focuses exclusively on the transformation of this quantity. Applying the beam-matrix transport relation yields
\begin{equation}
	\Sigma_{11}(n)
	= R_{11}^2(n)\,\Sigma_{11}(0)
	+ 2\,R_{11}(n)R_{12}(n)\,\Sigma_{12}(0)
	+ R_{12}^2(n)\,\Sigma_{22}(0) \; .
	\label{eq:matrix_measurement}
\end{equation}
This expression relates the measured beam size in the $n$-th configuration to the three unknown beam matrix elements in the reference state. The observable quantity is $\Sigma_{11}(n) = \sigma_n^2$, whereas the unknowns are the three independent beam matrix elements at the reference location. Since three unknown parameters have to be determined, at least three independent measurement conditions are required. Additional measurements provide redundancy, enabling consistency checks and improving the robustness of the~reconstruction through least-squares fitting.

Different measurement conditions can be realized in two ways. The first approach employs beam-profile measurements at different longitudinal locations, $n \mapsto s_n$, and is commonly referred to as the~\emph{multi-screen} method. The second approach keeps the measurement location fixed while varying the~beam optics, typically by changing the strength $k$ of a quadrupole magnet, $n \mapsto k_n$. This technique is known as the \emph{quadrupole-scan} method. Both approaches are discussed in more detail in the following.

\subparagraph{Multi-Screen Method}
The multi-screen method generates the required measurement conditions by recording beam profiles at several longitudinal positions $s$. For simplicity, consider a system with three profile monitors, which is sufficient to determine the three unknown beam matrix elements $\Sigma_{11}(0)$, $\Sigma_{12}(0)$, and $\Sigma_{22}(0)$ at a reference position $s_0$.

\begin{figure}[!t]
	\centering
	\includegraphics[width=0.86\textwidth]{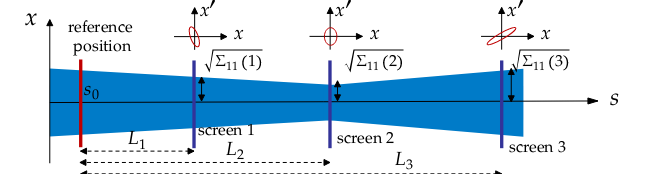}
	\caption{Three-screen method: profile measurements after known drifts determine the beam matrix at the reference point.}
	\label{fig:three_screen}
\end{figure}

The principle of the method is illustrated in \Fref{fig:three_screen}. Three profile monitors are installed in a~drift section downstream of $s_0$, at distances $L_n$ from the reference position, with $n=1,\ldots,3$. The~corresponding transport matrices from the reference position to the monitor locations are therefore given by the drift matrix of Eq.~(\ref{eq:drift}).

For this configuration, the resulting system of equations can be written as
\begin{align*}
	\Sigma_{11}(1) & = R_{11}^2(1)\,\Sigma_{11}(0) + 2\,R_{11}(1)R_{12}(1)\,\Sigma_{12}(0) + R_{12}^2(1)\,\Sigma_{22}(0) \\
	\Sigma_{11}(2) & = R_{11}^2(2)\,\Sigma_{11}(0) + 2\,R_{11}(2)R_{12}(2)\,\Sigma_{12}(0) + R_{12}^2(2)\,\Sigma_{22}(0) \\
	\Sigma_{11}(3) & = R_{11}^2(3)\,\Sigma_{11}(0) + 2\,R_{11}(3)R_{12}(3)\,\Sigma_{12}(0) + R_{12}^2(3)\,\Sigma_{22}(0),
\end{align*}
where $R(n)=R(s_0 \rightarrow s_n)$ is the transport matrix from the reference position $s_0$ to the $n$-th profile monitor. These equations can be transformed into the following compact matrix form:
\begin{equation*}
	\begin{pmatrix}\Sigma_{11}(1)\\\Sigma_{11}(2)\\\Sigma_{11}(3)\end{pmatrix}
	= \underbrace{
	\begin{pmatrix}
		R_{11}^2(1) & 2\,R_{11}(1)R_{12}(1) & R_{12}^2(1)\\
		R_{11}^2(2) & 2\,R_{11}(2)R_{12}(2) & R_{12}^2(2)\\
		R_{11}^2(3) & 2\,R_{11}(3)R_{12}(3) & R_{12}^2(3)
		\end{pmatrix}
	}_{\tilde{R}} \cdot
	\begin{pmatrix}\Sigma_{11}(0)\\\Sigma_{12}(0)\\\Sigma_{22}(0)\end{pmatrix} \; .
\end{equation*}
The vector of unknown beam-matrix elements at the reference location can then be obtained by inverting the matrix $\tilde{R}$,
\begin{equation*}
	\begin{pmatrix}\Sigma_{11}(0)\\\Sigma_{12}(0)\\\Sigma_{22}(0)\end{pmatrix} = \tilde{R}^{\mathsf{-1}} \cdot \begin{pmatrix}\Sigma_{11}(1)\\\Sigma_{11}(2)\\\Sigma_{11}(3)\end{pmatrix} \; .
\end{equation*}
Once the beam matrix elements at $s_0$ have been reconstructed, the emittance $\varepsilon$ follows directly from Eq.~(\ref{eq:rms_emit}). The corresponding Courant-Snyder (Twiss) parameters are then given by
\begin{equation*}
	\beta=\frac{\Sigma_{11}}{\varepsilon} \, ,
	\qquad
	\alpha=-\frac{\Sigma_{12}}{\varepsilon} \, ,
	\qquad
	\gamma=\frac{\Sigma_{22}}{\varepsilon} \; .
\end{equation*}

\begin{figure}[!t]
	\centering
	\includegraphics[width=0.86\textwidth]{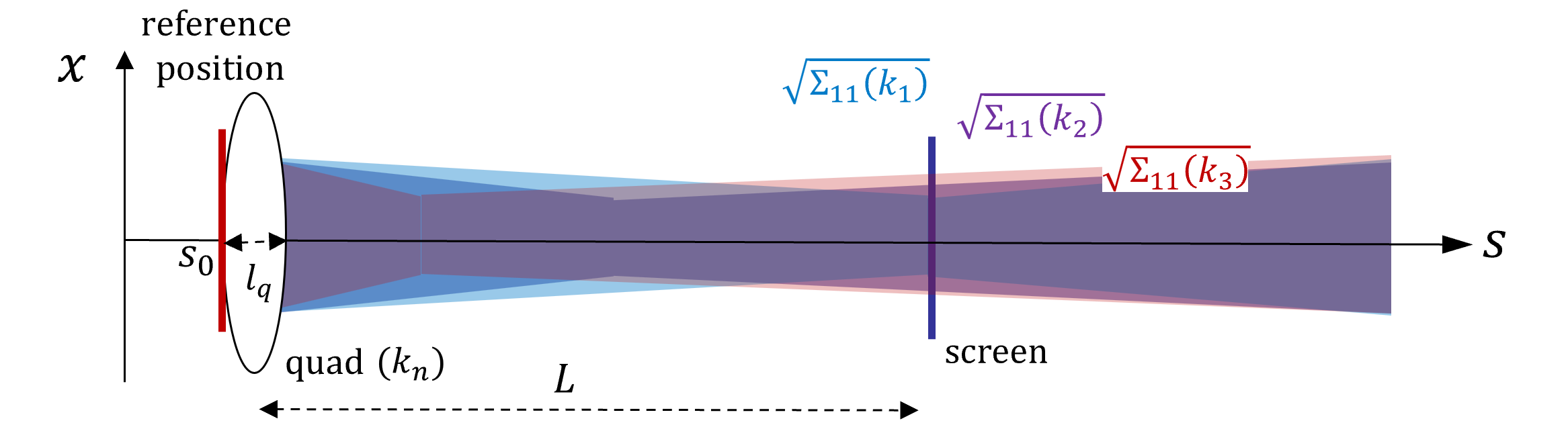}
	\caption{Single-quadrupole scan geometry.}
	\label{fig:quad_scan}
\end{figure}

\subparagraph{Quadrupole Scan}
In the quadrupole-scan method, beam profiles are measured at a fixed observation point, while different measurement conditions are generated by varying the focusing strength $k$ of an upstream quadrupole magnet. The principle of this technique is illustrated in \Fref{fig:quad_scan}. By changing the quadrupole strength in small steps, the beam experiences different focusing conditions, resulting in corresponding variations of the beam size at the observation point. These beam profiles are recorded with a single profile monitor located downstream of a drift section of length $L$. By convention, the reference position is chosen immediately upstream of the quadrupole magnet.

Compared to the multi-screen method, the principal advantage of a quadrupole scan is that the~number of measurement points is not limited by the number of available profile monitors. As a~result, a~large number of beam-size measurements can be acquired, improving the statistical accuracy and robustness of the fit. In principle, a similar increase in the number of data points could be achieved by installing additional profile monitors; however, this would require additional hardware and associated costs. The main disadvantage of the quadrupole-scan method is that it requires deliberate changes to the~beam optics. Consequently, such measurements generally cannot be performed parasitically and typically require dedicated machine time.

\begin{figure}[!t]
	\centering
	\includegraphics[width=0.55\textwidth]{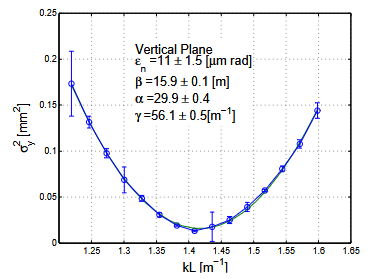}
	\caption{Example of a quadrupole-scan measurement. Figure reproduced from Ref.~\cite{Penco2008}.}
	\label{fig:quad_scan2}
\end{figure}

Using the beam-matrix transport relation, the matrix element $\Sigma_{11}$ associated with the beam size is again obtained as
\[
\Sigma_{11}(k_n, L)
= R_{11}^2(k_n, L)\,\Sigma_{11}(0)
+ 2\,R_{11}(k_n, L)R_{12}(k_n, L)\,\Sigma_{12}(0)
+ R_{12}^2(k_n, L)\,\Sigma_{22}(0) \; .
\]
In this case, the transport matrix is determined by both the quadrupole and the subsequent drift section. While the drift matrix $R_D$ is again given by Eq.~(\ref{eq:drift}), the quadrupole is treated in the thin-lens approximation, (i.e. focal length $f = 1/k\ell_q \gg \ell_q$ with $\ell_q$ the magnetic length). Within this approximation, the~quadrupole transfer matrix is given by
\[
R_Q=\begin{pmatrix}1&0\\K&1\end{pmatrix}
\qquad \mathrm{with} \quad K=k\ell_q  \quad \mathrm{the\ integrated\ quad\ strength,}
\]
and the total transport matrix is
\[
R=R_DR_Q=\begin{pmatrix}1+LK&L\\K&1\end{pmatrix}.
\]
Inserting the transport matrix into the beam-matrix transport relation results in a quadratic dependence of the matrix element $\Sigma_{11}$ on the integrated quadrupole strength $K$,
\begin{equation}
	\Sigma_{11}(K) = \underbrace{L^2\Sigma_{11}(0)}_{a} \cdot K^2 + \underbrace{2L\,(\Sigma_{11}(0) + L\Sigma_{12}(0))}_{b} \cdot K + \underbrace{\Sigma_{11}(0) + 2L\Sigma_{12}(0)+ L^2\Sigma_{22}(0)}_{c} \; .
	\label{eq:quad_scan}
\end{equation}
The procedure for an emittance measurement based on a quadrupole scan is therefore as follows. In a first step, the beam size $\sigma(k)=\sqrt{\Sigma_{11}(k)}$ is measured as a function of the quadrupole strength $k$. The measured beam sizes are then expressed in terms of $\sigma^2$ and plotted as a function of the integrated quadrupole strength $K=k\ell_q$. The resulting data are fitted with the parabolic function given by Eq.~(\ref{eq:quad_scan}). From the fit coefficients $a$, $b$, and $c$, the beam matrix elements $\Sigma_{11}(0)$, $\Sigma_{12}(0)$, and $\Sigma_{22}(0)$ at the reference location can be reconstructed. The beam emittance then follows directly from Eq.~(\ref{eq:rms_emit}). As an example, \Fref{fig:quad_scan2} shows a quadrupole scan performed at the ELETTRA injector \cite{Penco2008}. It should be noted that an~alternative parametrization of the quadrupole-scan parabola, based on Ref.~\cite{Ross1987}, is also commonly used in the literature. Further details can be found in the original reference.

While the parabolic dependence observed in a conventional quadrupole scan can be understood straightforwardly, modern FEL facilities often employ multi-quadrupole scan techniques, as introduced in Ref.~\cite{Prat2014}, rather than varying a single quadrupole treated as a thin lens. In this case, the analysis is no longer based on fitting Eq.~(\ref{eq:quad_scan}). Instead, the cost function used in the $\chi^2$ minimization is obtained directly from the beam-matrix transport relation for $\Sigma_{11}$, Eq.~(\ref{eq:matrix_measurement}).

The principal advantage of using several quadrupoles is the increased flexibility in the choice of optical settings. In particular, the optics can be adjusted such that the beam size at the profile monitor remains within a convenient range throughout the scan. This avoids excessively small beam sizes, which may be affected by space-charge-induced beam broadening and measurement resolution effects. Such conditions are especially critical near the minimum of the parabola, where the sensitivity of the~fit to the beam matrix parameters is greatest. Furthermore, an appropriate choice of optics settings in a~multi-quadrupole scan can provide sufficient information to reconstruct the beam matrices in both transverse planes simultaneously. Consequently, both transverse emittances can be determined from a single scan. In contrast, a conventional single-quadrupole scan generally requires separate measurements for the horizontal and vertical planes.

\subsection{Circular Accelerators}

While the Courant-Snyder (Twiss) parameters in a linac or transfer line are generally not known a priori and therefore have to be determined experimentally for emittance measurements, the situation is fundamentally different in a circular accelerator. In this case, one can exploit the periodicity of the beam optics, which requires the Twiss parameters to reproduce themselves after one revolution. According to Eq.~(\ref{eq:transferm}), this periodicity condition can be expressed mathematically as
\[
	\begin{pmatrix} \beta (s_0) \\ \alpha (s_0) \\ \gamma (s_0)\end{pmatrix} \stackrel{def}{=} \begin{pmatrix} \beta (s_0 + \mathcal{C}) \\ \alpha (s_0 + \mathcal{C}) \\ \gamma (s_0 + \mathcal{C})\end{pmatrix} = 
	\begin{pmatrix}
		R_{11}^2 & -2 R_{11} R_{12} & R_{12}^2 \\
		-R_{11} R_{21} & R_{11} R_{22} + R_{21} R_{12} & -R_{12} R_{22} \\
		R_{21}^2 & -2 R_{21} R_{22} & R_{22}^2
	\end{pmatrix}
	\cdot
	\begin{pmatrix} \beta (s_0) \\ \alpha (s_0) \\ \gamma (s_0) \end{pmatrix}
\]
with $\mathcal{C}$ the machine circumference. The solution of this Eigenwert problem yields \cite{Wille}
\begin{equation}
	\beta_0 = \frac{2\,R_{12}}{\sqrt{2 - R_{11}^2 - 2 R_{12}R_{21} - R_{22}^2}},\qquad
	\alpha_0 = \frac{R_{11} - R_{22}}{2\,R_{12}} \cdot \beta_0, \qquad
	\gamma_0 = \frac{1 + \alpha_0^2}{\beta_0} \; .
\end{equation}
Since the matrix elements $R_{ij}$ depend on the longitudinal position $s$ in the ring, the resulting values of the Courant-Snyder parameters are specific to the location at which the periodicity condition is evaluated, here the position of the profile monitor $s_0$. With the beta function known at this location, the emittance can be obtained directly from a single beam-size measurement $\sigma$ according to Eq.~(\ref{eq:Bsize}).

In practice, the beta function is often known with considerably higher precision than implied by the ideal lattice model. Accurate values can be obtained from particle-tracking simulations that account for field errors, magnet misalignments, nonlinearities, and other machine imperfections. Furthermore, the beta function can often be determined experimentally using methods such as $k$-modulation or phase-advance analysis; for a detailed discussion, see Ref.~\cite{MintyZimmermann2003}.

In contrast to linacs and transfer lines, the placement of profile monitors in circular accelerators is often constrained by boundary conditions, such as the requirement to extract synchrotron radiation for beam diagnostics. Consequently, the monitor signal is frequently generated at a location with finite dispersion, which must be taken into account in the emittance determination. Provided that the particle distributions are Gaussian and that the dispersion $D$ and relative momentum spread $\delta=\Delta p/p$ are known, the betatron and dispersive contributions to the beam size are statistically independent and therefore add in quadrature. Using Eq.~(\ref{eq:DispContrib}), the measured beam size is given by
\begin{equation}
	\sigma^2 = \beta\varepsilon + \left(D\frac{\Delta p}{p}\right)^2
	\label{eq:ring_size}
\end{equation}
from which the emittance $\varepsilon$ can be determined directly.

However, in modern ring-based fourth-generation light sources, the momentum spread $\delta$ is not necessarily constant and may vary with the machine conditions. As a consequence, the dispersive contribution to the measured beam size can change as well and must be taken into account when determining the emittance. In such cases, Eq.~(\ref{eq:ring_size}) can be exploited to determine both the transverse emittance and the momentum spread simultaneously. If beam-profile measurements are available at two locations with significantly different dispersion values, such that one measurement is primarily sensitive to the betatron contribution and the other to the dispersive contribution, the unknown quantities, $\varepsilon$ and $\delta$, can be extracted simultaneously according to
\begin{equation}
	\begin{pmatrix}
		\sigma_{x_1}^2 \\ \sigma_{x_2}^2 \\ \sigma_{y_1}^2 \\ \sigma_{y_2}^2
	\end{pmatrix}
	= \begin{pmatrix}
		\beta_{x_1} & 0 & D_{x_1}^2 \\ 
		\beta_{x_2} & 0 & D_{x_2}^2 \\ 
		0 & \beta_{y_1} & D_{y_1}^2 \\ 
		0 & \beta_{y_2} & D_{y_2}^2
	\end{pmatrix}
	\begin{pmatrix}
		\varepsilon_{x} \\ \varepsilon_{y} \\ \delta^2
	\end{pmatrix},
	\label{eq:emittance_dispersion}
\end{equation}
where $\sigma_{x_{1,2}}$ and $\sigma_{y_{1,2}}$ are the measured horizontal and vertical beam sizes at the two observation points, $\beta_{x_{1,2}}$ and $\beta_{y_{1,2}}$ denote the corresponding beta functions, and $D_{x_{1,2}}$ and $D_{y_{1,2}}$ represent the horizontal and vertical dispersion values at the respective monitor locations.

\subsection{FEL Slice Emittance} 

In high-gain Free-Electron Lasers (FELs), the FEL interaction takes place locally within the electron bunch and is therefore governed primarily by the properties of individual longitudinal slices rather than by the projected parameters of the bunch as a whole. Due to collective effects such as coherent synchrotron radiation and space-charge forces, the beam properties may vary significantly along the~bunch. The \emph{slice emittance} denotes the transverse emittance of a narrow longitudinal section (or slice) of the~bunch and serves as a measure of the local beam quality. It is a key parameter for FEL performance, as it directly characterizes the overlap between the electron beam and the radiation field, and consequently the achievable gain. For this reason, there is a strong demand for diagnostic techniques capable of resolving beam parameters on the temporal scale of the bunch length.

\begin{figure}[!t]
	\centering
	\includegraphics[width=0.7\textwidth]{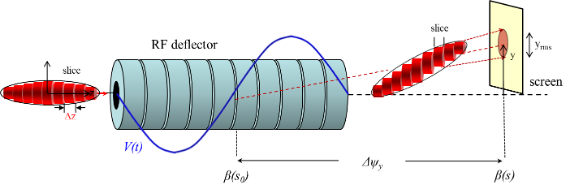}
	\caption{Working principle of a Transverse Deflecting Structure (TDS). A bunch traversing the cavity at the RF zero-crossing receives a time-dependent vertical ($y$) kick that is zero at the bunch center and increases linearly towards the head and tail of the bunch. Downstream of the TDS, the vertical position of the electrons is therefore correlated with their longitudinal coordinate. A fast kicker magnet (not shown) directs the bunch onto an off-axis screen for measurement.}
	\label{fig:tds}
\end{figure}

A powerful technique for this purpose is the use of a radio-frequency transverse deflecting structure (TDS), which imparts a time-dependent transverse kick to a single electron bunch, as illustrated in \Fref{fig:tds}. More detailed descriptions of the operating principle can be found in Refs.~\cite{Kube2026,Levefre25}. Downstream of the TDS, the transverse particle distribution is measured with a screen monitor (usually a scintillator or OTR). In the resulting image, the coordinate along the streaking direction is mapped onto the longitudinal position within the bunch, thereby providing direct access to time-resolved beam properties.
\begin{figure}[!b]
	\centering
	\includegraphics[width=0.9\textwidth,clip]{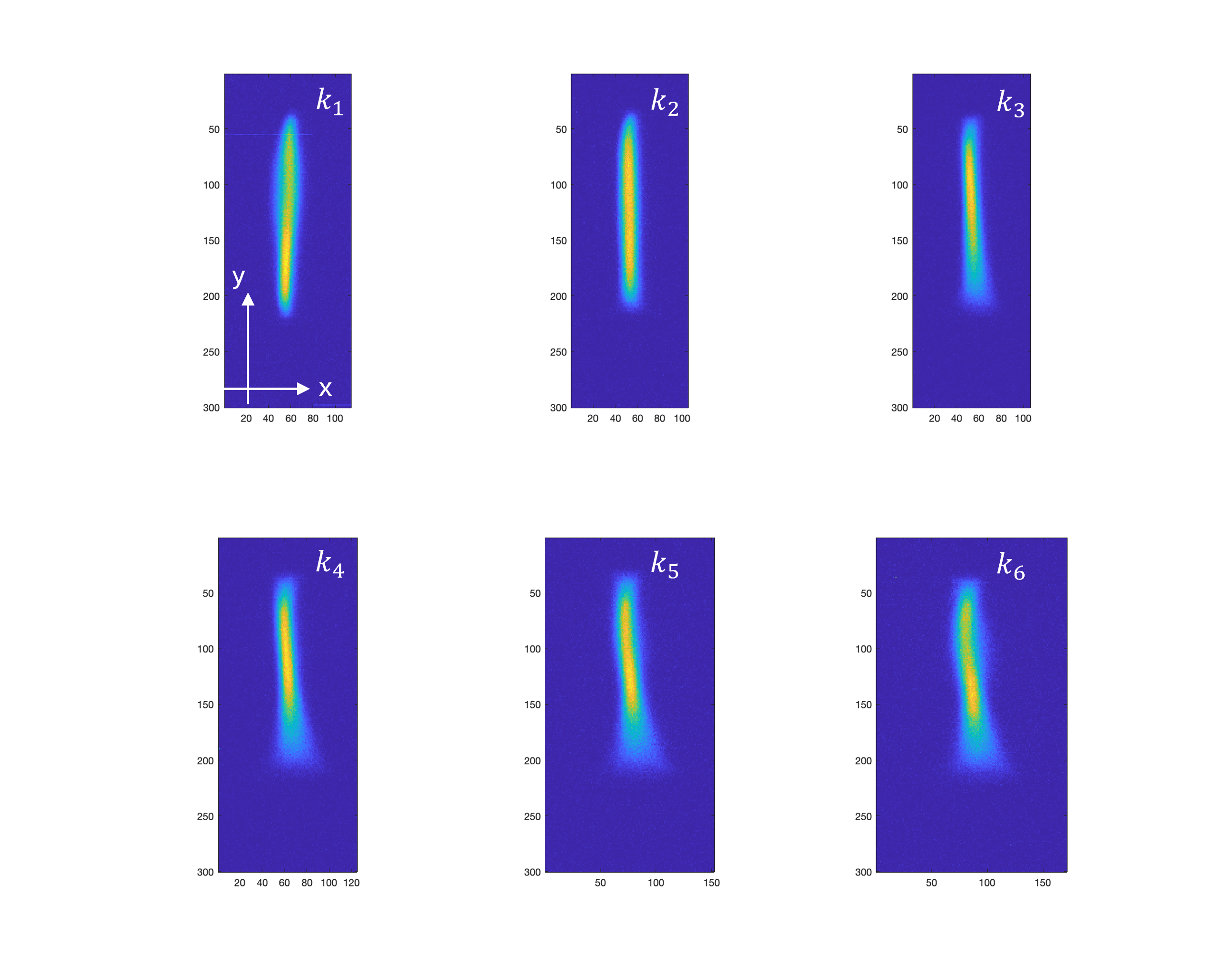}
	\vspace*{-.8cm}
	\caption{Example of a slice-emittance analysis. The streaked beam image is subdivided into longitudinal slices, and for each slice the transverse beam size is evaluated for a series of quadrupole settings $k_i$ ($i=1,\ldots,6$). Figures adapted from the European XFEL Logbook.}
	\label{fig:slice1}
\end{figure}
For illustration, \Fref{fig:slice1} shows examples of streaked beam images measured at the European XFEL. The~bunch was streaked in the vertical ($y$) direction, such that the vertical coordinate corresponds to the longitudinal position within the bunch. Consequently, the $y$-axis contains the time-resolved beam information, whereas the transverse plane orthogonal to the streaking direction preserves the transverse (in the present case the horizontal) beam-profile information ($x$). By subdividing the image into narrow slices along the~vertical direction, the horizontal beam profile can be analyzed independently for each longitudinal slice of the bunch.

As described in Section~\ref{sec:BMschemes}, beam-matrix-based emittance measurement techniques can then be applied to each slice individually. In this way, the slice emittance can be reconstructed as a function of the~longitudinal position (slice index) within the bunch. In the European XFEL example shown in \Fref{fig:slice1}, this was achieved by performing a quadrupole scan, as indicated by the different quadrupole strengths $k_i$. For each quadrupole setting, a streaked beam image was recorded and analyzed slice by slice, yielding the beam-size dependence required for the emittance reconstruction.
\begin{figure}[!t]
	\centering
	\includegraphics[width=0.55\textwidth]{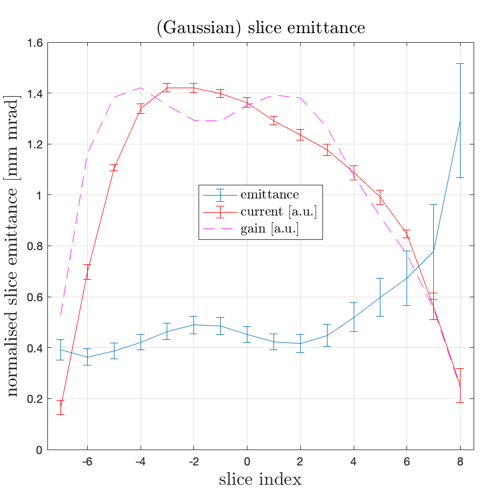}
	\caption{Reconstructed slice emittance, current profile, and FEL gain obtained from the quadrupole-scan measurement shown in \Fref{fig:slice1}. Figure adapted from the European XFEL Logbook.}
	\label{fig:slice2}
\end{figure}
The final outcome of the analysis is shown in \Fref{fig:slice2}, which presents the reconstructed slice emittance, the bunch current profile, and the~corresponding FEL gain derived from the quadrupole-scan measurement of \Fref{fig:slice1}.

\section*{Acknowledgements}
I would like to thank my colleagues Bolko Beutner and Matthias Scholz from DESY for their valuable input and the many fruitful discussions on emittance diagnostics. I am also grateful to all colleagues from the beam diagnostics community who kindly provided presentation material and figures that could be used in the preparation of this lecture.


\begin{thebibliography}{99}
\bibitem{CourantSnyder1958}
E.D. Courant and H.S. Snyder,
\emph{Theory of the alternating-gradient synchrotron}, Annals of Physics {\bf 3} (1958) 1.
\href{https://doi.org/10.1016/0003-4916(58)90012-5}
{10.1016/0003-4916(58)90012-5}

\bibitem{Wiedemann2015}
H. Wiedemann, \emph{Particle Accelerator Physics}, 4th ed, (Springer Cham, 2015).\\
\href{https://doi.org/10.1007/978-3-540-49045-6}
{10.1007/978-3-540-49045-6}

\bibitem{MintyZimmermann2003}
M.G. Minty and F. Zimmermann,
\emph{Measurement and Control of Charged Particle Beams},
(Springer, Berlin, 2003).
\href{https://doi.org/10.1007/978-3-662-08581-3}
{10.1007/978-3-662-08581-3}

\bibitem{Schmuser2008}
P. Schmüser, M. Dohlus, J. Rossbach, and Ch. Behrens,
\emph{Ultraviolet and Soft X-Ray Free-Electron Lasers: Introduction to Physical Principles, Experimental Results, Technological Challenges}, 
(Springer Cham, 2014).
\href{https://doi.org/10.1007/978-3-319-04081-3}
{10.1007/978-3-319-04081-3}

\bibitem{Liouville1838}
J. Liouville, \emph{Note sur la théorie de la variation des constantes arbitraires}, Journal de Mathématiques Pures et Appliquées {\bf 3} (1838) 342.
\href{http://eudml.org/doc/234417}
{<http://eudml.org/doc/234417>}

\bibitem{Wilson2001}
E.J.N. Wilson, \emph{An Introduction to Particle Accelerators}, (Oxford University Press, Oxdord, 2001).
\href{https://doi.org/10.1093/acprof:oso/9780198508298.001.0001}
{10.1093/acprof:oso/9780198508298.001.0001}

\bibitem{Forck2024}
P. Forck, \emph{Beam Instrumentation}, Proceedings of the Joint Universities Accelerator School (JUAS) 2024, CERN-2024-003.
\href{https://doi.org/10.23730/CYRSP-2024-003.1339}
{10.23730/CYRSP-2024-003.1339}

\bibitem{Torino2026}
L. Torino, \emph{Transverse Profile Measurements}, these proceedings.

\bibitem{Kube2026}
G. Kube, \emph{Measurement Principles}, these proceedings.

\bibitem{Alesini2007}
D. Alesini \emph{et al.}, \emph{Experimental results with the SPARC emittance-meter}, Proc. PAC'07, Albuquerque (New Mexico, USA), June 2007, MOOAAB02, p.80 (2007).

\bibitem{Cianchi2008}
A. Cianchi \emph{et al.}, \emph{High brightness electron beam emittance evolution measurements in an rf photoinjector}, Phys. Rev. ST Accel. Beams \textbf{11} (2008) 032801.
\href{https://doi.org/10.1103/PhysRevSTAB.11.032801}
{10.1103/PhysRevSTAB.11.032801}

\bibitem{Zhang1996}
M. Zhang, \emph{Emittance Formula for Slits	and Pepper-pot Measurement}, FERMILAB-TM-1988 (1996).

\bibitem{Penco2008}
G. Penco \emph{et al.}, \emph{Beam Emittance Measurement for the New Full Energy Injector at ELETTRA}, Proc. EPAC'08, Genoa (Italy), June 2008, TUPC079, p.1236 (2008).

\bibitem{Ross1987}
M.C. Ross \emph{et al.}, \emph{Automated Emittance Measurements in the SLC}, Proc. 1987 IEEE PAC, Washington D.C. (USA), March 1987, p.725 (1987).

\bibitem{Prat2014}
E. Prat and M. Aiba, \emph{Four-dimensional transverse beam matrix measurement using the multiple-quadrupole scan technique}, Phys. Rev. ST Accel. Beams \textbf{17} (2014) 052801.\\
\href{https://doi.org/10.1103/PhysRevSTAB.17.052801}
{10.1103/PhysRevSTAB.17.052801}

\bibitem{Wille}
K. Wille, \emph{The Physics of Particle Accelerators}, (Oxford University Press, Oxdord, 2009).

\bibitem{Levefre25}
T. Lef\`evre, \emph{Short bunch length measurements}, these proceedings.

\end{thebibliography}
\end{document}